\documentclass[vecphys]{svmult}
\usepackage{makeidx}         
\usepackage{graphicx}        
\usepackage{multicol}         
\usepackage[bottom]{footmisc}

\makeindex             
\begin{document}
\title*{Top physics at the Tevatron collider}
\author{Fabrizio Margaroli \inst{1}}
\institute{Purdue University, \\
Fermilab PO Box 500 MS 312 \\
60510 Batavia, Illinois, USA \\
e-mail: \texttt{margarol@fnal.gov}}
\maketitle
The top quark has been discovered in 1995 at the CDF\,\cite{CDF} and D\O\,\cite{D0} experiments located in the Tevatron ring at the Fermilab laboratory. After more than a decade the Tevatron collider, with its center-of-mass energy collisions of 1.96\,TeV, is still the only machine capable of producing such exceptionally heavy particle. Here I present a selection of the most recent CDF and D\O\ measurements performed analyzing $\sim 1$\,fb$^{-1}$ of integrated luminosity ${\cal L}$.
\section{Introduction}
\label{sec:intro}
The top quark is produced at the Tevatron mostly in pairs: the theoretical cross section for this process amounts to 6.7\,pb at Next-to-Leading Order (NLO)\,pb\,\cite{Cacciari} (for an assumed top quark mass of 175\,GeV/c$^2$). According to the Standard Model (SM), the top quark decays in a W boson and a b quark $\simeq 100\%$ of the time. The hadronic or leptonic decays of the two W bosons thus characterize the three non-overlapping final samples, which differ for their Branching Ratios (BR) and background contamination and composition: the {\bf dileptonic} sample, with two leptonically decaying W's, has two high-Pt tracks and large missing E$_T$, and is the cleanest of all, but on the other hand the one with the smallest BR($\sim$5$\%$). The {\bf semileptonic} sample contains events where one W decays leptonically and the other decays hadronically; it is characterized by large BR($\sim$30$\%$) and moderate background, mostly coming from production of W in association with jets.  The {\bf all-hadronic} sample is where both W's decay hadronically; this channel has the largest BR($\sim$44$\%$) but also a very large background from QCD multijet production. In the latter two cases, to enhance signal purity CDF and D\O\ require the presence of long-lived B mesons, as a signature of b quarks, through the presence of a displaced secondary vertex (b-tagging).
\section{Top quark properties}
Thanks to high statistics and high purity, semileptonic $t \bar t$ events are the best candidates to test SM predictions and non-SM particle production in the top sector:  
\paragraph{Pair production cross section}
The measurement of the $t \bar t$ production cross section provides a test of QCD calculation and any discrepancy from the theoretical expectation could hint to production or decay mechanisms not predicted by the SM. The most recent measurement comes from the D\O\ experiment\,\cite{D0xsec} and is performed in the semileptonic channel using ${\cal L} =1$\,fb$^{-1}$, counting events passing selection cuts and requiring at least one jet to be tagged as a b-quark jet; the measured cross section corresponds to $\sigma_{t \bar t} = 8.3^{+0.6}_{-0.5}(stat.)^{+0.9}_{-1.0}(syst.)\pm0.5(lumi.)\,$pb. The measurements performed by CDF and D\O\ in the complementary samples give compatible results.
\paragraph{Production mechanism}
SM predicts the top pairs to be produced through quark-antiquark annihilation 85\% of the time, and the rest 15$\%$ through gluon-gluon fusion. Taking advantage of the fact that the average number of low-$P_T$ tracks is proportional to the gluon content of a sample, CDF deploys a template method to fit a gluon-rich and a gluon-deprived track multiplicity distribution to the data, to measure\,\cite{CDFprod} the fraction of events produced through gluon-gluon fusion to be 
$\sigma(gg \to t \bar t)/\sigma(p \bar p \to t \bar t) = 0.07\pm0.14(stat.)\pm 0.07(syst.)$.
\paragraph{Decay mechanism}
According to the SM the W boson is produced 70$\%$ of the time with longitudinal helicity, and the rest with left-handed helicity.; right handed helicity is forbidden by the theory. A template method is used here, the template variable being $cos \theta^*$, the cosine of the decay angle between the momentum of the charged lepton in the W boson rest frame and the W momentum in the top quark rest frame, which is highly sensitive to the W helicity. CDF measures\,\cite{CDFwhel}
$F^0 = 0.59 \pm 0.12 (stat.)^{+0.07}_{-0.06} (syst.)$ and $F^+ = -0.03 \pm 0.06 (stat.)^{+0.04}_{-0.03} (syst.)$.
\paragraph{New physics with top quarks?}
The top quark can be seen as an hadronic probe to very high mass scales. 
CDF scans the $t \bar t$ invariant mass distribution to look for possible peaks due to resonant Z$^{\prime}$ production in the mass range 450-900\,GeV/c$^2$. Limits can be set to the product of the cross section times the branching ratio to top pairs. This limit amounts\,\cite{CDFresonance} to $\sigma \times$ BR(Z$^{\prime} \to t \bar t) < 0.8$\,pb at 95$\%$ Confidence Level (CL) for a Z$^{\prime}$ mass greater than 600\,GeV/c$^2$. \\

Overall, the measurements performed by the two experiments are in good agreement with each other and with the theoretical prediction.
\section{The top quark mass}
\label{sec:mass}
The top quark is the only quark that decays before hadronizing. Its mass, which is a free parameter in the SM, can thus be direcly measured. Moreover, due to the top quark and W contribution to radiative corrections, the measurements of their masses provide a powerful constraint on the Higgs boson mass. The top quark mass has traditionally been measured in each channel; a major boost in precision has been achieved by exploiting the presence of hadronically decaying W whose daughter jets can be used to constrain the biggest source of systematic uncertainty, the Jet Energy Scale (JES). For this reason, the most precise results now come from the analysis of the semileptonic and the all-hadronic samples.
There are two main classes of methods to extract the mass: the Template Method and the Matrix Element method. 
The former consists in choosing a variable which is strongly correlated with the observable one wants to measure, and in building templates of this variable for simulated signal and background events. The variable used to measure the M$_{\rm top}$ is a tri-jet reconstructed invariant mass; the light quark dijet mass is chosen to simultaneously measure the JES. 
The Matrix Element technique aims to use all the available informations to calculate a probability for the event to come from signal or background according to the theory predictions for the final state kinematics. Transfer functions are needed in order to convert reconstructed objects into kinematical tree-level quantities.  For both techniques a likelihood will compare the data to the signal and background and its maximization will provide us the measured values. 
The most precise measurements are performed using the matrix element technique in the semileptonic sample to simultaneously measure M$_{\rm top}$ and JES. The most recent D\O\ measurement\,\cite{D0mass} amount to M$_{\rm top} = 170.5 \pm 1.8(stat.) \pm 1.6 ($JES$) \pm 1.2 (syst.)\,$GeV/c$^2$. 
\begin{figure}
\centering
\includegraphics[height=5.0cm]{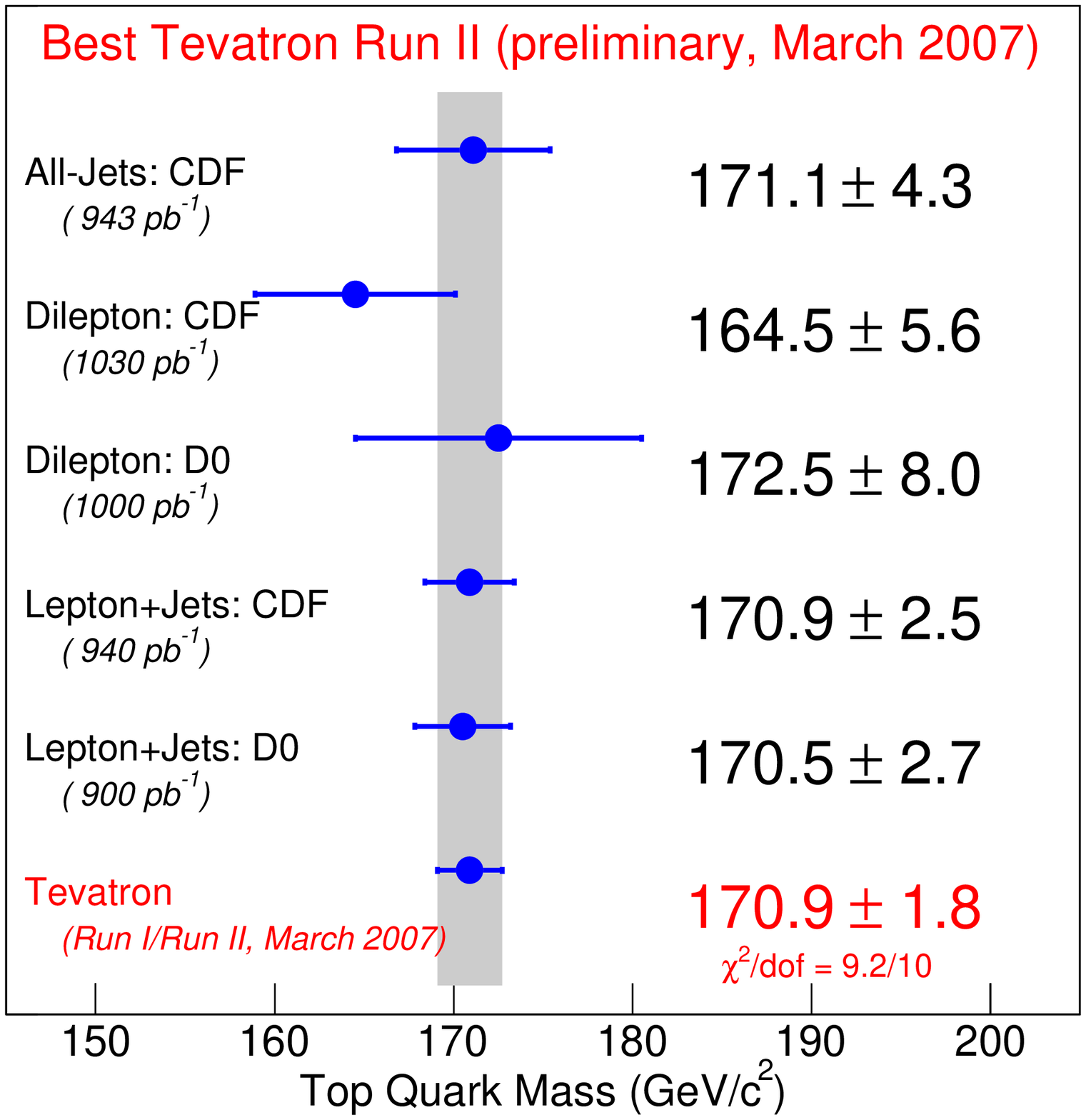}
\includegraphics[height=5.4cm]{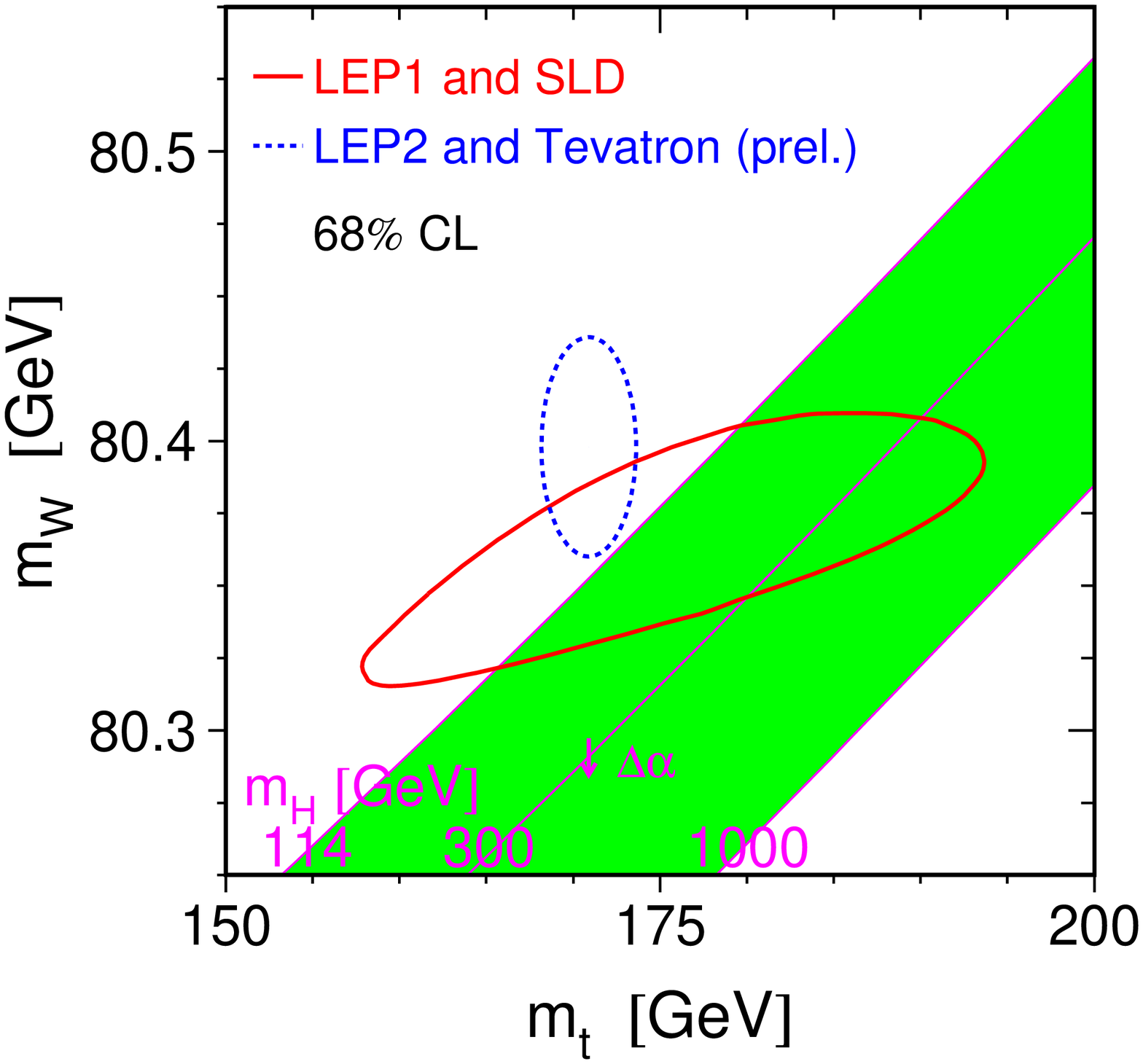}
\caption{Tevatron Run II best measurements used in the combination; on the right is shown the updated constraint on the SM Higgs mass given from the latest determination of the top and W masses.}
\label{fig:1}       
\end{figure}
CDF alone explores the all-hadronic channel, where the latest analysis employs a cut based selection to improve the signal-to-(mostly QCD) background ratio from $\sim1/1000$ to $\sim1/1$. This analysis uses a mixed technique to extract the mass: a template is built out of the probability given by the matrix element computation, and a dijet mass is used to measure the JES; this result\,\cite{CDFhadmass} is now the most precise in this channel and corresponds to M$_{\rm top} = 171.1 \pm 2.8(stat.) \pm 2.4($JES$) \pm 2.1 (syst.)\, {\rm GeV/c}^2$.
The best measurement in each channel is then combined to give the very precise Tevatron average value\,\cite{TevTopMass} of M$_{\rm top} = 170.9 \pm 1.1 (stat.) \pm 1.5 (syst.) = 170.9 \pm 1.8$\, GeV/c$^2$.
With such a $1 \%$ precision achieved, the M$_{\rm top}$ measurement will likely be a long-standing legacy of the Tevatron collider.
\section{Single top production}
The SM allows the electroweak production of single top quarks with the theoretical cross section at NLO\,\cite{Kidonakis} of 1.98\,pb in the t-channel and 0.88\,pb in the s-channel (assuming M$_{\rm top}=175\,$GeV/c$^2$). Single top quark events can be used to study the $Wtb$ coupling and directly measure the $V_{tb}$ element of the CKM matrix without assuming only three generation of quarks. CDF and D\O\ restrict their searches to events where the W decays leptonically; the signature is thus characterized by missing energy from the neutrino, one high-Pt lepton and, a b-jet from the top decay, which is required to be tagged to further reduce the background. Additionally we expect a light quark jet in the t-channel or one more b-jet in the s-channel. After the event selection we are left with a S/B of about 1/20. Both CDF and D\O\ experiments use different advanced techniques to better isolate the signal from the large background. The best D\O\ measurement uses a machine-learning technique that applies cuts iteratively to classify events, namely a boosted decision tree. It produces an output variable distribution which ranges from 0 to 1, with the background peaking close to 0 and the signal close to 1. A binned likelihood fit is used to extract the cross section, that D\O\ measures\,\cite{singletopD0} to be $\sigma($s+t channel$)=4.3^{+1.8}_{-1.4}\,$pb, $3.4\,\sigma$ away from the background only hypothesis, and in agreement with the SM expectation; D\O\ also measures the element V$_{tb}$ of the CKM matrix to be $0.68 < |V_{tb}|< 1$ at 95\% CL.
\begin{figure}
\centering
\includegraphics[height=4.1cm]{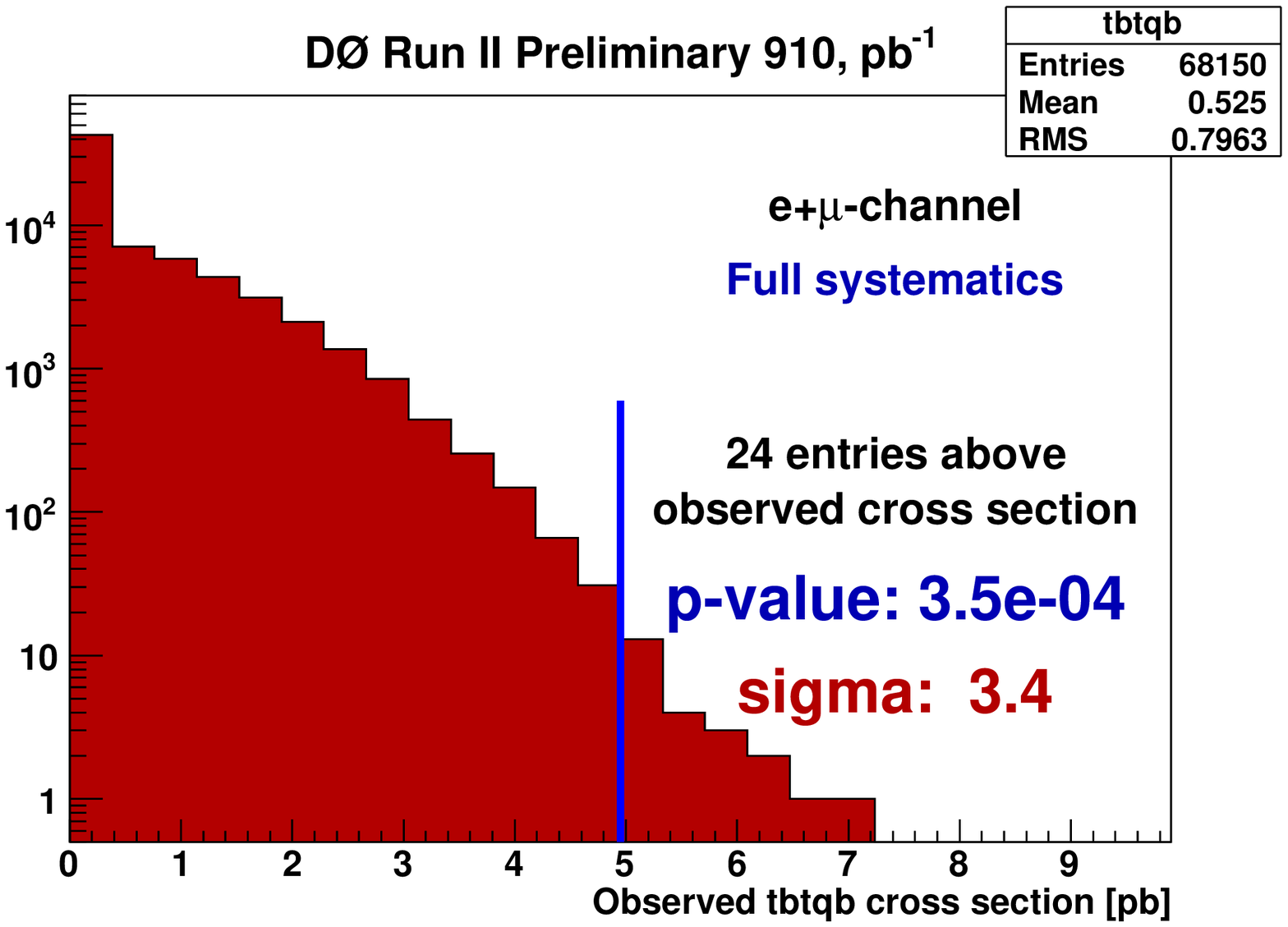}
\includegraphics[height=4.3cm]{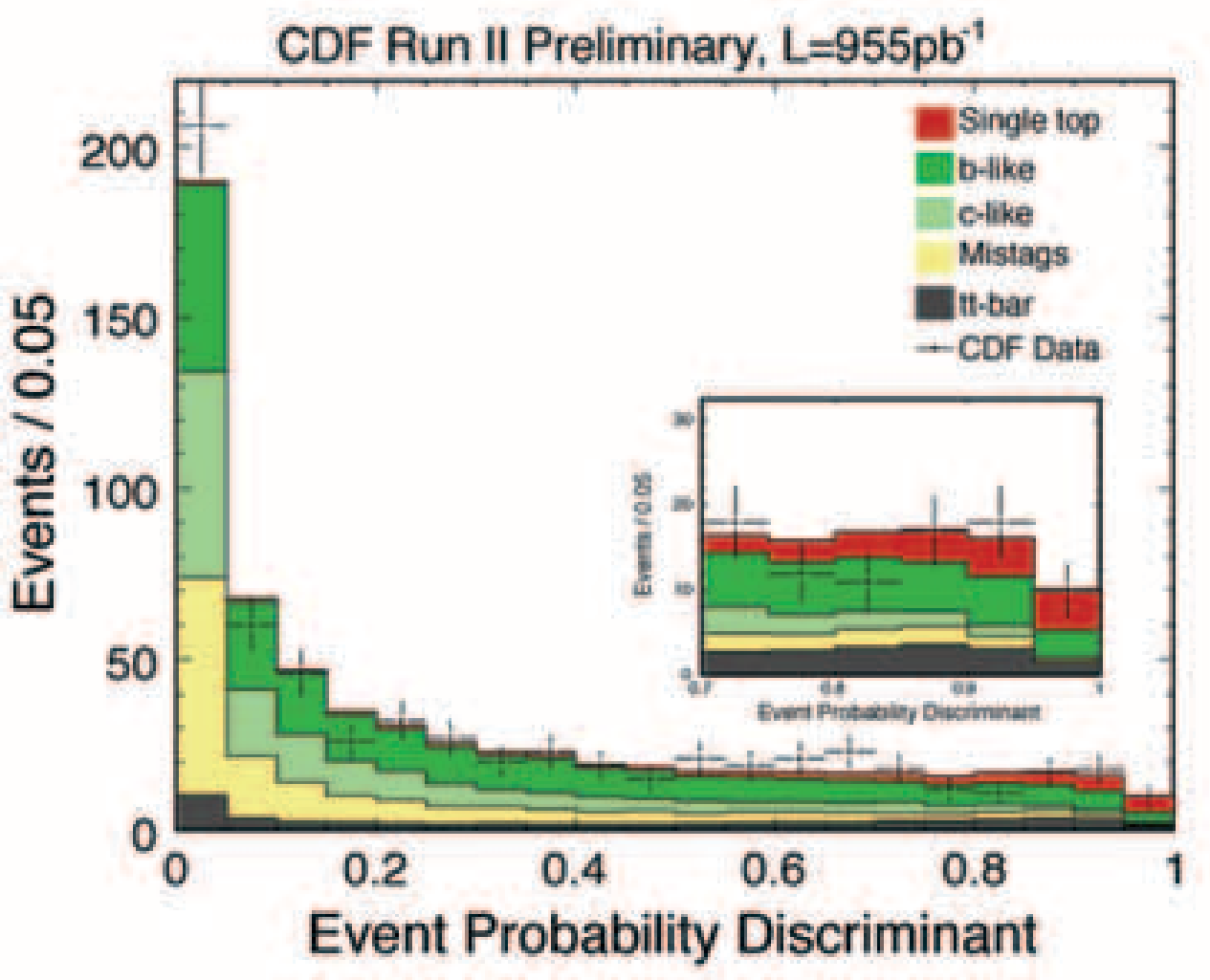}
\caption{On left, fraction of background-only pseudo-experiments giving a cross section higher than the observed. On the right, event probability discriminant used by CDF to extract the cross-section.}
\label{fig:1}       
\end{figure}
CDF's best result comes from using the event matrix element to build a probability for the event to come from signal or background. An event probability discriminant is then built and a likelihood fit extracts the signal and background relative normalizations. CDF measures\,\cite{singletopCDF} an excess of $2.3\,
\sigma$ and extract a cross section for the s+t channel to be $2.7^{+1.5}_{-1.3}\,$pb.
\section{Conclusions}
The measurement presented here confirm the SM expectation for top quark production and decay within the theoretical uncertainty, and provide high precision on the most important top property, the top quark mass, that it will take years for the LHC to achieve it. The first evidence of single top production and first direct measurement of the $V_{tb}$ parameter constitute another major Tevatron success. However, most analysis are statistically limited and with 2\,fb$^{-1}$ already recorded, and between 6 to 8\,fb$^{-1}$ expected, uncertainties will be reduced and smaller deviation from the SM investigated. 
I would like to thank here the conference organizers and my CDF and D\O\ collaborators for the hard work and effort spent in achieving the results presented above.
%
%
%

%
%

%
\printindex
\end{document}